\documentclass[a4paper,12pt]{report}
\usepackage[utf8]{inputenc}
\usepackage{graphicx}
\usepackage{subfigure}
\usepackage{hyperref}
\usepackage{cleveref}
\usepackage{cite}
\usepackage{setspace}
\usepackage[nottoc,notlof,notlot]{tocbibind}

\begin{document}
\doublespacing
\author{Ebtesam Almazrouei, Raed M. Shubair, and Fabrice Saffre }
\title{ Internet of NanoThings:\\ Concepts and Applications}
\date{2018}

\maketitle
\tableofcontents

\chapter{Internet of NanoThings}
This chapter focuses on Internet of Things from the nanoscale point of view.  The chapter starts with section 1 which provides an introduction of nanothings and nanotechnologies. The nanoscale communication paradigms and the different approaches are discussed for nanodevices development. Nanodevice characteristics are discussed and the architecture of wireless nanodevices are outlined.
Section 2 describes Internet of NanoThing(IoNT), its network architecture, and the challenges of nanoscale communication which is essential for enabling IoNT.
 Section 3 gives some practical applications of IoNT. The internet of Bio-NanoThing (IoBNT) and relevant biomedical applications are discussed. Other Applications such as military, industrial, and environmental applications are also outlined.

\section{Introduction to Nanonetworks }

\subsubsection{General Introduction }
In 1959, the Nobel physicist Richard Feynman stated the important role of tiny atoms and molecules size to develop fully functional an advanced nano devices. The challenge of scaling the metrial in the atoms nano scale was highlighted. Also, he entitled that the research in engineering  will be focused how to redesign or create nanocomponents in nanoscale devices.  Nowadays, current technologies facing the challenge of how to develop these nanocomponents with taking into consideration its nanoscale phenomena \cite{ akyildiz2011nanonetworks}.  

The term nanotechnology was first introduced by N. taniguchi in 1974 as follows: “Nanotechnology mainly
consists of the processing of, separation, consolidation, and deformation of materials by one atom or by one molecule.” \cite{taniguchi1974basic, akyildiz2011nanonetworks}.  In 1986, K. Eric Drexler associated the basic idea of Feynman’s vision about nanodevices and added the concept of ability of creating these nanocomponents by replicating themselves using computer control instead of billion tiny factories  controlled by human\cite{ akyildiz2011nanonetworks,drexler1987molecular}. However, the attention from researcher towards nanoscale components has slowly increase until the advancement  starts to show only in the early of 2000s.

\subsection{Nanotechnology }
Many people could have confused between the nanoscience and nanotechnology. Nano technology is  distinguished as applying the technology to create novel materials at nano dimentional scale based on the knowledge from the nanoscience. The nanoscience is defined as investigating the properties of the material and it’s phenomena at the nano scale. The  research interest   of the knowledge in nanoscienece and nanotechnology is increased worldwide. This enables exploiting new advances materials, devices, and technologies to work in few nano-meter length \cite{roszek2005nanotechnology}. 
Nanotechnology is an emerging technology providing a new sets to create and control the structure of the engineering materials at nanoscale dimension for the aggregates of each individual molecules. This will enable the nano scale components  to perform defined tasks  such as data storing, sensing, computing, and actuation \cite{akyildizoptoelectronics}.  All the nano components will be integrated in a single advanced nanodevice. where this device will be able to achieve complex task in a distributed manner for health care, military, biological, and nanosensor network \cite{ akyildiz2010electromagnetic,akyildizoptoelectronics,shubair_robust_2005, shubair_performance_2005, belhoul_modelling_2003, shubair_robust_2004, al-ardi_direction_2006, nwalozie_simple_2013, al-nuaimi_direction_2005, bakhar_eigen_2009,mohjazi2011deployment}.  

\subsection{Nanoscale Communication Paradigms }
The communication capabilities of nanodevices plays vital role in the nanotechnology. It is important to enable accurate synchronous between nanodevices that work in a cooperative and supervised environments. The communication in nano devices could be divided two two categories as follows: 
\begin{itemize}
\item Internal nanocommunication: Communication between two or more of nanodevices.
\item External nanocommunication: Communication between the nanodevice and external system such as another electronic micro device.

\end{itemize}

There are different communication technologies have been proposed in the literature for nanodevices such as electromagnetic communication, molecular communication, and acuastic communication \cite{ akyildiz2008nanonetworks, freitas1999nanomedicine}.

\textbf{\textit{Electromagnetic communication}} uses electromagnetic waves which propagates via air or wire with less losses. It is well used in microelectronics devices, but it has some limitations in implementing electromagnetic  connection through wires in nanoscale devices. Therefore, the electromagnetic communication for nanodevices should be implemented through wireless communication. The wireless connection require nanoscale antenna to be developed for nanodevices, also a radiofrequency transceiver should be implemented in the nanodevice to establish bidirectional electromagnetic wireless communication \cite{ akyildiz2008nanonetworks, freitas1999nanomedicine}. However, the integration of current radiotransciver is challenging due to the complexity and the limitation of the size in nanodevices. Also, the insufficient output power of nanotransiver affect the establishment of bidirectional connection between nanodevices. Hence, the electromagnetic waves could be used to send information from microdevice to nanomechine in one direction only. Therefore, another communication technology should be used to enable the internal communication among nanodevices and the external communication from a nanodevice to a microdevice \cite{ akyildiz2008nanonetworks, freitas1999nanomedicine,khan_compact_2017, shubair_closed-form_1993, omar_uwb_2016,alhajri2018accurate,samhan2006design, alhajri2015hybrid}.  
 
\textbf{\textit{Acoustic Communication}} is based on the   transmission of ultrasonic waves. The ultrasonic transducers are integrated in the nanodevices, therefore there are able to sense the rapid variations of the pressure coming from ultrasonic waves; then acoustic signals are emitted \cite{ akyildiz2008nanonetworks}.

 \textbf{\textit{Nanomechanical communication}} provides the ability to transmit the information through hard junctions between linked nanodevices. This communication paradigm requires a physical link between transmitter and receiver in the nanodevices.  In addition, the desired mechanical transceivers should be aligned precisely  which is main drawback of this communication paradigm as the nanodevices will be deployed in nanonetwork without any direct or physical contact between them. Moreover, a precise navigation systems is needed to locate the nanodevices in order to establish correct nanomechanical communication \cite{ akyildiz2008nanonetworks}.

\textbf{\textit{Molecular communication}} is a new and promising technology that enables transmit and receive the data in molecules \cite{shubair2015vivo, elayan2017terahertz, elayan2017wireless, elayan2016channel, elayan2016vivo, elayan2017bio,moore2006design, elayan2018end, elayan2017photothermal, akyildiz2008nanonetworks,elayan2017multi, elayan2018vivo, alnabooda2017terahertz}. The natural environment and the size of the molecules make it is more feasible to integrate the  molecular transceivers in nanodevices. The nanotranscievers are capable to  release some molecules, react to others , and response to internak commands between molecules in the nanonetwork.

\subsection{Development of Nanodevices}

\begin{figure}[h!]
        \centering
        \includegraphics[width = 13cm,height = 7cm]{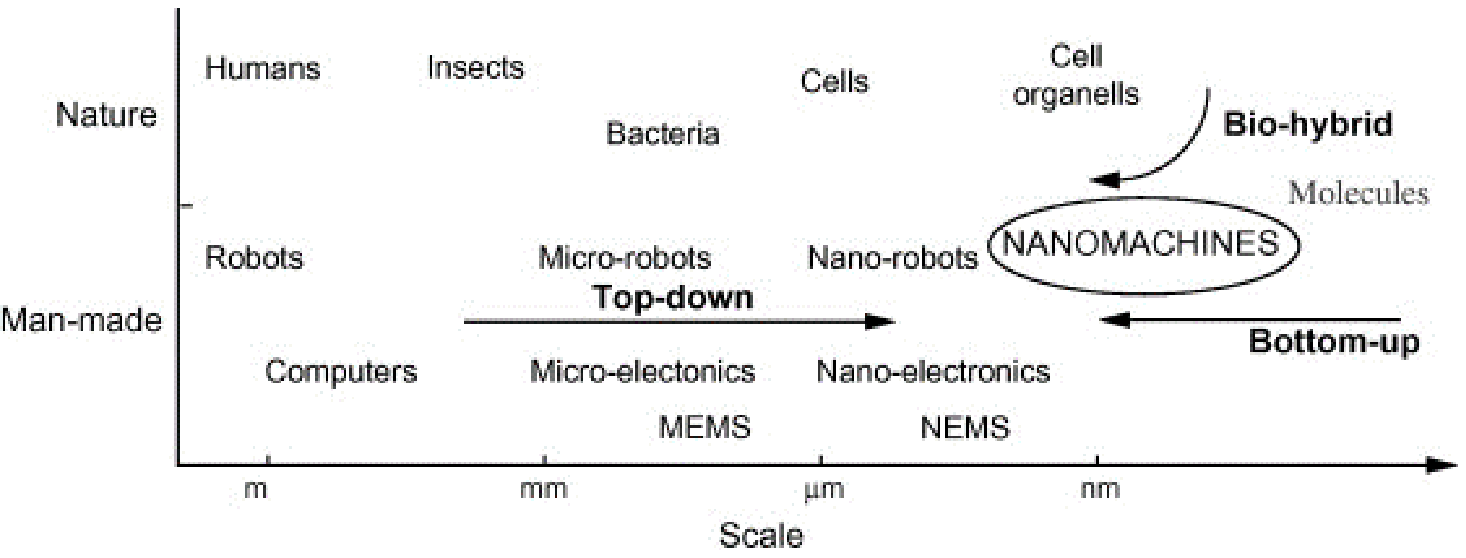}
         \caption{Nanodevice approaches\cite{akyildiz2008nanonetworks}.}
        \label{fig:nanodevicearch1}
       \end{figure}

A nanosensordevice is defined as a device designed from nanocomponents to perform in nanoscale and able to establish a required task such as communicate, sense, compute, store data, and actuate. These tasks performed by nanocomponents and the complexity of the developed-nanodevice relies on the level of the requested task. Different approaches are used to develop the nanodevice: 1) the topdown approach, the bottom-up approach, and the bio-hybrid approach as depicted in Figure \ref{fig:nanodevicearch1}.

\subsubsection{Top-down approach} 
In the topdown approach, nanodevices are developed by downscaling the current microelectronic and micro-electro-mechanical technologies. Advanced  manufacturing processes are used to develop the nanodevice such as electron beam lithography \cite{tseng2003electron,burek2009fabrication, emami2015electron} and  micro-contact printing \cite{lee2004large}. The architecture of nanodevices are kept same as the architecture of the  microelectronic devices and micro-electro-mechanical systems (MEMS). However, the nano-electromechanical systems  approach \cite{chang2003highlights,goldstein2005race,meyyappan2006nanotechnology}. Simple mechanical structures fabrication method such as nano-gears is proposed for the fabrication and assembly of these nanodevice following this approach but is still at an early stage \cite{yun2007manipulation,soe2015nanogears}
 
 \subsubsection{Bottom-up approach} 
 In the bottom-up approach, building blocks such as individual molecules are used to develop nanodevices. Nanodevices has been theoretically designed such as  molecular differential gears and pumps [51] based on a discrete number of molecules \cite{peterson2000taking}. This approach is based on molecular manufacturing technologies where nanodevices are assembled molecule  by molecule. This technology is not available yet \cite{drexler1992nanosystems,wolf2015nanophysics,akyildiz2008nanonetworks}. Currently there are different methods used to develop the nanodevices following bottom-up approach based on self-assembly molecular properties \cite{balzani2006artificial} such as molecular switches \cite{ballardini2001artificial} and molecular shuttles \cite{balzani1998molecular}.

\subsubsection{Bio-hybrid approach} 

Bio-hybrid approach proposed using the existing biological nanodevices such as molecular motors, as building blocks or models for the development of the new nanodevices \cite{whitesides2001once}. Most of the biological living organisms are exist in cells as shown in Figure \ref{fig:nanodevicearch1}. The feature of the biological structure of the living cell such as nano-biosensors, nanoactuators, biological data storing components, tools and control units is expected to form new baseline of the manmade-nanodevice \cite{eric1991unbounding,akyildiz2008nanonetworks}. 

 Several biological nanodevices are interconnected and form nanonetwork. The inter-cell communication technique allows multiple cells to cooperate to perform complex tasks such as cell division, the control  of hormonal activities or immune system responses in  humans. This operation of this biological nanonetwork is based on molecular signaling.
 
  The optimized architecture, power consumption, and communication paradigm of the existing biological nanodevices motivates the development of the future nanodevices using bio-hyprid approach.

\subsection{Nanodevice Characteristics}

The development of the future nanodevices relies on the advanced technology that able to design the future nanodevices which will be available in the near future. The main characteristics of the future nanodevices are detailed as follows \cite{akyildiz2008nanonetworks}:

\begin{itemize}
        
\item Self-contained nanodevice: Each nanodevice has a code or set of instructions to realize the intended task. The code or the instruction set can be embedded in the molecular structure of nanodevice or the nanodevice can read it from neighboured-molecular structure.

\item Self-assembly nanodevices: Nanodevice can form an organized structure from several disordered elements without external intervention, using the local interactions between them. self-assembly feature is naturally found in molecular affinities between two different elements at nano level. This process will enable the nanodevice to interact in an autonomous way with external molecules.

\item Self-replication: This characteristic should be included in the nanodevice to enable the nanodevice to copy itself using external elements. It implies that the nanodevice has set of instructions to create a copy of itself. This feature will facilitate the ability to realize macroscopic tasks by creating large number of nanodevices in an inexpensive way \cite{merkle1992self}. 

\item Nanodevice-Communication: The communication between nanodevices is crucial to allow nanodevices to cooperate with each other in order to accomplish or realize more complex tasks.

\item Locomotion: Nanodevice moves from one place to another by spatial-temporal actuation. Locomotion will help the nanodevices to perform specific tasks by identifying the nanodevice location to be at the right place and the right time to conduct the intended task. However, the single nanodevice cannot move towards a previously identified target. Embedded nanosensors and nanopropellers could be used in a complex system to detect and trace the target location. This characteristics will optimize the use of nanorobots for disease treatments in healthcare \cite{cavalcanti2007nanorobot,muthukumaran2015role}.

Further advances in nano-sensors and nano-actuators are expected to enable the integration
of molecular transceivers into nano-machines.

\end{itemize}

\subsection{Wireless Nanodevice Architectures}

A wireless nanodevice could consists of one or more components based on the level of complexity required to perform intended task. The architecture of nanodevice whether it is nanorobot or simple molecular switches is as follows \cite{akyildiz2008nanonetworks,cavalcanti2007nanorobot}:

\begin{enumerate}
        \item  Control unit: Executes the instructions to perform the requested tasks through controlling all the other components of the nanodevice. Also, the information of the nanodevice could be saved in a storage unit inside the control unit.
        \item  Communication unit: Enables the transmission and the reception of the massages at nanoscale device e.g., molecules through nanotransceivers.
    \item  Reproduction unit: Utilizes external elements to fabricate each nanodevice's component  and assemble all the components to replicate the nanodevice. All the instructions to conduct this task are installed in the unit. 
    \item Power unit: Provides the power to the entire components in the nanodevice, harvests energy from external sources,e.g., temperature, and light. The energy is stored for a future needs of distribution and consumption.
        \item Sensor and actuators : Plays a role as interface components between the nanodevice and the environment and the nano-machine. There are various types of sensors or actuators can be enclosed in a nanodevice design such as chemical sensors, temperature sensors, pumps, clamps, motor or locomotion mechanisms.
        
\end{enumerate}

\section{The Internet of NanoThings (IoNT)}
\subsection{Internet of Nano-Things} 

The Internet of Things (IoT) has gained a lot of interest from researchers in the last decade. The objective is to extend the internet to many devices and objects from different domains by interconnecting those objects and devices with embedded computing capabilities \cite{akyildiz2015internet}. The word “things” includes all the physical object on the planet 
not only communication devices to be connected to the Internet, and controlled through wireless networks \cite{kawamoto2014internet}. IOT devices will interconnect through various types of short-range wireless technologies such as WiFi, radio frequency identification (RFID), ZigBee, and sensor networks \cite{feki2013guest,miraz2015review}. 

The concept of IoT has attracted  many researchers worldwide. It covers many areas such as body area networks, home area networks, Unmanned Aerial Vehicle (UAV) networks, Device-to-Device (D2D) communications, and satellite networks. Different networking protocols, applications and  network domains is expected to be integrated  to fit IoT technologies in the near future \cite{kawamoto2014internet}. Security features and management protocols are expected to be added to IOT linked networks and devices \cite{miraz2015review}.

\begin{figure}[h!]
        \centering
        \includegraphics[width = 13cm,height = 13cm,angle=270]{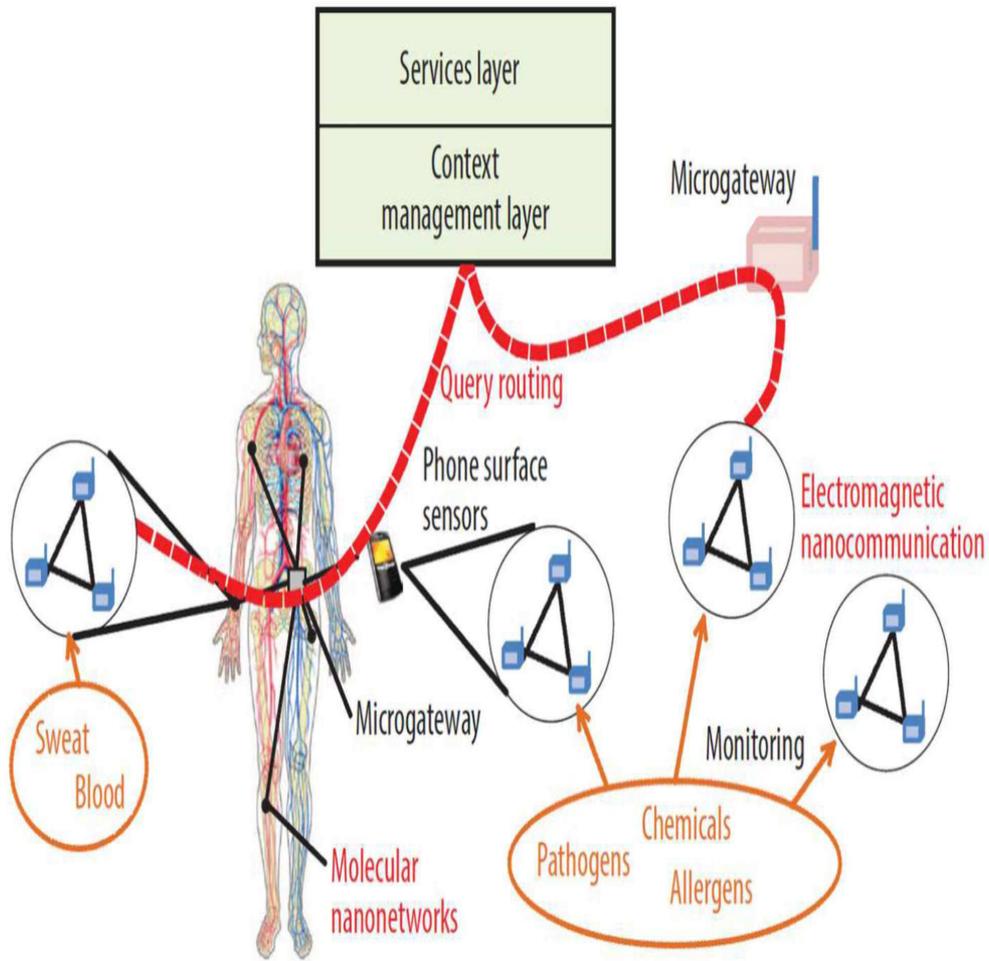}
         \caption{The Internet of Nano Things \cite{balasubramaniam2013realizing}.}
        \label{fig:nanobodynetworktex}
       
\end{figure}

In IoT,all types of real physical elements such as actuators and sensors, personal  or home electronic devices are connected among others which enable new era of seamless connectivity for various applications such as machine to machine (device to device) communication, real time monitoring for health care and industrial environment, vehicle to vehicle communication and transportation, smart grids  and infrastructures to establish smart energy management, infrastructure management, environmental monitoring, intelligent health monitoring, intelligent transportation on large grid \cite{akyildiz2015internet}. This is achieved by incorporating nanodevices  to be interconnected using nanonetworks. Figure \ref{fig:nanobodynetworktex} illustrates the concept of nanonetworks in healthcare application.

Implementing nanodevices facilitates the ability to sense and collect data
data from places in the body that previously impossible to sense due to sensor size. Thus, new medical diagnostics and discoveries will advance the current medical technology \cite{balasubramaniam2013realizing}. Akyildiz et al. define this technology as Internet of Nano Thing (IoNT) \cite{akyildiz2010internet}. 

The concept of the IoNT is introduced as a type of IoT where nanodevices whose dimensions may range from 1 to 100 nm are interconnected with classical networks leading to new networking paradigms. Graphene-based nanoantennae  is proposed to be utilized in IoNT technology and operating at Terahertz frequency band \cite{akyildiz2010internet}. The problem of extreme attenuation of tetrahertz frequencies in nanoscale device  is outlined by \cite{akyildiz2010internet}. IoNT faces challenges to interface the current microdevices networking with the new nanodevices scheme. Therefore, major research should be conducted to address the communication and networking challanges in electromagnetic field, the channel modelling, and the required networking protocols to operate in IoNT for various industrial, biomedical and industrial applications.

\subsection{Network Architecture} 
\begin{figure}[h!]

        \centering
        \includegraphics[width = 13cm,height = 7cm]{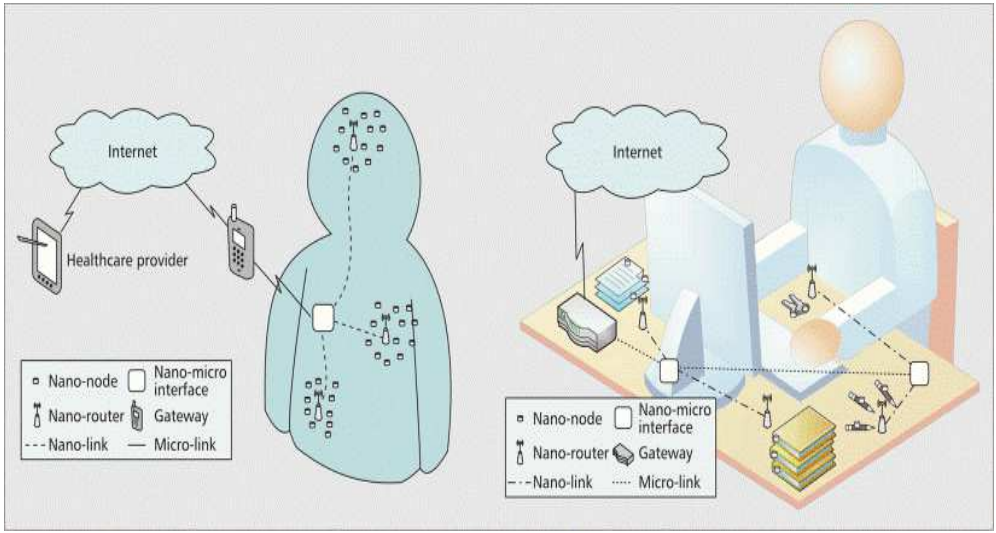}
         \caption{Network architecture for the Internet of Nano-Things \cite{akyildiz2010internet}.}
        \label{fig:networkarchalkyl}
       
\end{figure}

The work conducted by Akyildiz et al. focuses  on electromagnetic communication for the IoNT networks \cite{akyildiz2010internet} in 
intrabody nanonetworks for remote healthcare, and the interconnected office. The network architecture shows in Figure \ref{fig:networkarchalkyl} is composed of nano-nodes, nano-routers, nano machines  such as nanosensors and nanoactuators which deployed in the human body to provide  the examiner or the healthcare provider the ability to access and control the nanodevices remotely through nanomicro interface devices \cite{akyildiz2010electromagnetic}. Additionally, Akyildiz et al. shows the interconnected office architecture where each single element found in the office is provided with nanotransciever to allow them to stay permanently connected to the internet. Therefore, the  location and the status of the all elements in the office are tracked in an effortless manner. However,  an ultra-low power consumption and reasonable computing capabilities are required  for the nanodevices to harvest the mechanical and electromagnetic energy from the environment and keep the function with high performance \cite{wang2008towards,akyildiz2010electromagnetic}.

However, each nanonetwork has fundamentals components in the network
architecture of the Internet of Nano-Things as following \cite{akyildiz2010internet}:
\subsubsection{Nano Nodes}
Nanonodes are the smallest and simplest nanomachines in the nanonetwork architecture. Simple computation tasks are assigned to them because of their limited memory. There is limitation in their communication capabilities and consumption energy, therefore they are able to transmit for short range distances. Nanonodes could be implemented in all types of things such as books, keys, paper folders, or inside the human body as biological nanosensor nodes.

\subsubsection{Nano Routers}
Nanorouters are capable for aggregating the information coming from limited nanomachines. It is used to control the behavior of nanonodes by exchanging simple controlling commands such as: on/off switching, read value, sleep, etc.). Nanorouters have larger computational resources than nanonodes, however, the increase in capabilities leads to increase in their size which affect the their deployment in the nanontwork.
\subsubsection{Nano-micro Interface Devices}
Nano-micro interface devices are implemented to enable the receive/send information coming from nanorouters to the microscale device or system and vice versa. Nano-micro interface devices could be hybrid devices able to communicate in the nanoscale using the classical communication paradigms in conventional communication networks and new communication paradigm for nanodevices network.

\subsubsection{Gateway}
Gateway facilitates the remote connection of the entire system over the Internet. For example, in an intrabody network scenario in healthcare era, an advanced cellphone can forward the received information from nano-micro interface device to the end user ( healthcare provider in this example). A modem-router is utilized in interconnected office to establish this functionality.
it receives from a nano-micro interface in our
wrist to our healthcare provider. In the interconnected
office, a modem-router can provided this
functionality.
Despite the interconnection of microscale
devices, the development of gateways and the
network management over the Internet are still
open research areas, in the remaining of this
article we mainly focus on the communication
challenges among nanomachines.

Moreover, the work by Balasubramaniam et al. focuses
on wireless body area networks constructed by nano
devices [2]. The body area networks collect vital patient information
and feed those information to service providers’
computing systems. As a consequence, it achieves higher
accuracy and efficiency in monitoring the health conditions
of a large number of patients. Moreover, sensors embedded
in the environment can passively assist daily life of the elderly
and disabled people. With the development of small
devices and their communications performance, such networks
in tiny area are also expected to be required in the
future.

\subsection{ Challenges of Nanoscale Communication} 

The Internet of Nano-Things requires redesign and develop new communication paradigms, and networking concepts that will be compatible for nanoscale machines. Many communication challenges appear in the physical layer of nanomachines to the nanonetworking protocols. This section highlights the main challenges from communication respective as discussed in \cite{akyildiz2010internet}.

\subsubsection{Frequency Band of Operation of Electromagnatic Nano-transceivers}
The communication opportunities and challenges at the nanoscale devices are strongly associated with the operating frequency band of the nano-transciver especially nano-antennas. Theses future antennas are predicted to be manufactured using noval material as graphene for nano communication network \cite{akyildiz2010internet, jornet2010graphene, da2009carbon}.
The velocity of wave propagation in grahene-nanoantenna is scaled to be one hundred times below the speed of light in vacuum. Additionally, the resonant frequency nanoantenna  built with graphene can be up to two orders of magnitude below that  nanoantennas based ob non-carbon materials.

In particular, Lin et al. found that a 1 $\mu$m long  graphene-based nano-antenna built either by graphene nanoribbon (GNR) or carbon nanotube
(CNT) radiates efficiently in Terahertz range which satisfies the predication of frequency band for graphene-based RF transistors \cite{lin2010100}. In \cite{jensen2007nanotube}, it has been shown that a single carbon nanotube  that mechanically resonates at the wave frequency is able to receive and demodulate an electromagnetic wave. This single CNT antenna  has designed  with one end connected to very high voltage source and the other end is left floating. The electrons at the free tip end vibrate when the nanotube is irradiated by an EM wave. Thus, there will be frequency initiated by EM wave and if it matches the natural resonant frequency of the CNT antenna, these vibrations become significant and enable the single CNT antenna to receive and demodulate the signal.

The EM waves generated by CNT-based nano-mechanical receiver can operated above few micrometers. However, the energy efficiency is predicted to be very low to generate EM waves in nano antennas radiated in Terahertz band \cite{weldon2008nanomechanical}. Also, a high power source is required to excite the CNT antenna mechanically which is inefficient for generating the future EM nanonetworks operating in the Terahertz band. Nevertheless, CNT-based nano-mechanical receiver can be used in the nanonetwork  to control the nanodevices from the macro and microscale in nano-micro interface devices. As an example, a conventional AM/FM transmitter can be used to activate/deactivate thousands of nanodevices simultaneously.

The far infrared band and the microwaves frequency band which is above and below, respectively the Tetrahertz band have been extensively investigated. The Terahertz band is one he least-explored frequency zones in the EM spectrum. Therefore, the new channel models  for the Terahertz band should  be developed for electromagnetic nanonetworks.

\subsubsection{Channel Modeling}
The Terahertz band spans the frequencies between 100 GHz and 10 THz is still unlicensed band. It has major limitations for short and medium range communications \cite{piesiewicz2007short, jornet2010channel} but but it is applicable for nanonetwork applications as discussed aforementioned, therefore the channel modelling for this band in the very short range should be investigated. Jornet et al. investigates the properties of the Terahertz band in terms of path-loss, noise, bandwidth and channel capacity as described below \cite{jornet2010channel}. 

\begin{itemize}
\item\textbf{Path-loss}

The total path-loss $L(f_{w}, l_{path})$ for a travelling wave in the Terahertz band is defined as the addition of the spreading loss $L_{spread}$ and the molecular absorption loss $L_{absorbtion}$. 

\begin{equation}
L(f_{w}, l_{path}) = L_{spread} (f_{w}, l_{path}) + L_{absorbtion} (f_{w}, l_{path})
\end{equation}
where $f_{w}$ is the frequency wave in Tetrahertz band and $l_{path}$ is the total path length of the wave.

The spreading loss is a result of the attenuation coming from the expansion of the frequency wave $f_{w}$ as it propagates through the medium $l_{path}$.

\begin{equation}
L_{spread} (f_{w}, l_{path})=(\frac{4  \pi f_{w} l_{path}}{c})^{2}
\end{equation}
where $c$ is the speed of light in a vacuum.

The absorption loss $L_{absorbtion}$ is a result of the attenuation occurs because of the molecular absorption that affects the propagation wave. The wave energy converts to kinetic energy of the excited molecules by electromagnetic radiation at certain frequencies within the terahertz band where part of the radiation  converts to internal vibration. Thus, the wave energy reduced leading to the absorption loss $L_{absorbtion}$ and it defined as follows:

\begin{equation}
L_{absorbtion} (f_{w},l_{path}) =\frac{1}{e^{-k(f_{w}) l_{path}}}
\end{equation}
where $k$ is the medium absorption coefficient.

The absorption loss depends on the type of the molecules and its concentration along the path. Different resonance frequency associated to different types of molecules  where the absorption at each resonance spreads
over a range of frequencies. As a consequence, the Terahertz channel will suffer from high frequency selectivity, multi-path propagation, and scattering from the nano particles in the field which affect the  signal
strength at the receiver.

\item\textbf{Noise}

The main source of the ambient noise in the Terahertz band is the molecular noise. The molecular absorption introduces noise along with the attenuation. This type of noise occurs only when transmitting signal through the channel. Additionally, equivalent noise temperature is introduced around the frequencies where the molecular absorption is considered high. The total noise power at the receiver is computed as follows:

\begin{equation}
P_{noise} (f_{w},l_{path}) = k_{B} B (T_{molecular} (f_{w},l_{path}) + T_{else} (f_{w}))
\end{equation}
where $k_{B}$ refers to the Boltzmann constant, $B$ is the system transmission bandwidth, $T_{molecular}$ is the the molecular noise temperature, and $T_{else}$ stands for other noise source present
in the medium, e.g., electronic noise of the receiver.

The total noise power $P_{noise}$ has several peaks of noise in the spectrum due to the different resonant frequencies associated with each type of molecules \cite{jornet2010channel}.

\item\textbf{Bandwidth and channel capacity}

The molecular absorption determines the  transmission bandwidth in Terahertz channel. Therefore, the molecular composition of the medium and the total transmission path constrain the available bandwidth. The available bandwidth  for a very short range is ranging from a few hundreds of gigahertz to almost ten Terahertz (almost the entire band). 

Therefore, the channel capacity of electromagnetic nanonetworks in the Terahertz band is predicted to be in the order of a few terabits per second. However, there is limitation in the transmitted information capacity due to limitation in the capabilities of nanomachines or nanodevices which does not make use of this large bandwidth. Despite of this limitation, the available bandwidth could open new research for new information modulation techniques and channel sharing schemes, specially designed for nanodevice in the nanonetworks operating in the Terahertz band.

\end{itemize}
\section{ Applications of Internet of NanoThings (IoNT)}

\subsection{The Internet of Bio-NanoThings (IoBNT)} 

A noval research directed towards implementing nanodevices and nanotechnology in the biological field. There is increased interest to merge the tools from synthetic biology within the nanotechnology to control, modify, reengineer, and reuse the biological cells \cite{kahl2013survey, akyildiz2015internet}. The biological cell which is utilized in IoT embedded computing device is called Bio-NanoThing (BNT) where it can effectively control, reuse, and reengineer the functionalities of biological cells such as sensing, actuation, processing, and communication. This concept introduced the Internet of Bio-NanoThing (IoBNT) where the cells are based on biological molecules instead of electronics.

\subsubsection{Biomedical Applications}
IoBNT enables compatibility and stability at the bio-molecular level. This provides the ability to use IoNBT to interact with organs and tissues. In this section, IoBNT applications are mentioned \cite{akyildiz2008nanonetworks}.

\begin{itemize}
\item\textbf{Immune system support}

IoBNT can be utilized to support the immune system  to identify and control foreign and pathogen elements in the human body. Several nanodevices such as sensors and actuators collaborate with each other in macro, micro, and nano systems to protect organism against diseases. Implementing nanodevices can advanced the medical field by utilizing theses nanodevices to predict, detect, and eliminate certain procedures based on  localization of malicious agents and cells, such as cancer cells \cite{chen2005development,freitas2005nanotechnology}. This will minimize the risk of developing such disease and provide treatments less aggressive and invasive compared to the existing ones.

\item\textbf{Bio-hybrid implants}
Nanonetowrks in IOBNT will support the replacement of organs, nervous tracks, or lost tissues in the human body \cite{drexler1992nanosystems, freitas2005nanomedicine}. Friendly interfaces can be provided between the bio-hybrid implants and the environment which enable the restoration of central nervous system tracks. 

\item\textbf{Drug delivery systems}
Nanpdevices in IoBNT can be used as regulater implants that could compensate metabolism diseases such as diabetes.Smart glucose reservoirs and nanosensors collaborate to support the glucose level  mechanisms \cite{patra2013intelligent,freitas2006pharmacytes}. The effects of neurodegenerative diseases can be eleminated using drug delivery system to deliver neurotransmitters or specific drugs to neurosystem\cite{wowk1988cell}.

\item\textbf{ Health monitoring}
Implemeting nanosensor networks in human body can benefit the medical field by provide health monitoring to monitor and control Oxygen, cholesterol level, and  hormonal disorders, and provide early diagnoses of the health status \cite{freitas2005nanomedicine,donaldson198824th}. A good level of connectivity should be maintained between the nanonetwork and the actors who can access the transferred health information.

\item\textbf{ Genetic engineering}
The use of nanonetworks in IoNBT will allow  the potential increase of genetic engineering applications. Nanodevices will enables the modification, re-engineering, and manipulation of nano-structures inside genes and molecular sequences \cite{akyildiz2008nanonetworks}.

\end{itemize}

\subsection{Other Applications of IoNT} 

\subsubsection{Industrial Applications}
Nanonetwork will be used in the industrial and consumer goods applications. It will advance the manufacturing processes, the development of new materials, and the quality control procedures. More specifically, these applications
have already been proposed by \cite{akyildiz2015internet}:
\begin{itemize}
        \item\textbf {Food and water quality control}
        Nanonetworks could be used to monitor and control the food and fluids quality. Nanosensor will be able to detect the toxic components and small bacteria found in the food and water that can not be detected using traditional sensing technologies \cite{aylott2003optical}. This advanced self-powered nanosensor networks will be able even to sense the tiny amount of defects such as chemical or biological agents installed in water supplies \cite{akyildiz2015internet}.
        
        \item\textbf {Functionalized materials and fabrics}
                New advanced materials and fabrics can be manufactured by using nanonetworks in order to improve certain functionalities. There are developed products such as antimicrobial and stain-repeller textiles using nanofunctionalized materials \cite{ravindra2010fabrication,tessier2005antimicrobial}. Nanoactuators communicate with nanosensors in order to control the reaction which will improve the airflow in advanced smart fabric.
        \end{itemize}
        
         \subsubsection{Military Applications}
        
        Nanothechnology can emphasize and advance several  applications for military field. Nanonetwork range is short; therefore the range of nanonetworks is specified based on the required application. The range of nanonetworks for monitoring soldier performance applications is small within human body range while for a dense large network is required for battlefield
        monitoring and actuation. Below are some of the military applications: 
    \begin{itemize}
    	
        \item\textbf {Nuclear, biological, and chemical (NBC) defenses} 
        
        For large area over the battlefield or targeted areas, a dense network consists of nanosensors and nanoactuators is deployed to detect aggressive chemical and biological agents. Additionally, it coordinates the defensive response battlefield areas \cite{smalley2003carbon,akyildiz2008nanonetworks}. Nanosensor networks can be used be to detect the unauthorized entrance biological, chemical, and radiological materials installed  in the cargo containers.
        
        \item\textbf {Nano-functionalized equipments}
        Nanonetworks can advanced the camouflage and army uniforms using new advanced military equipments than can manufacture advanced materials equipeed with nanonetworks. This technology will enable the self-regulation of soldiers's body temprature underneath his clothes and will be able to detect and inform if the the soldier has been injured \cite{endo2006development}.
\end{itemize}
        
        \subsubsection{ Environmental Applications}
        Nanoneworks have various application in environmental fields that will advance the current technologies. Some environmental applications are mentioned as follows \cite{akyildiz2008nanonetworks}:
\begin{itemize}
        \item\textbf {Biodegradation}
        The rising problem of garbage handling around the world, biodegradation process in the garbage dumps using nanonetworks could manage the problem. Nano networks can be used to sense and tag different materials, then smart nanoactuators are utilized to locate and process the biodegradation for theses materials. 
        \item\textbf {Animals and biodiversity control}
        Several animal species can be controlled by nanonetworks in natural environments. Nanonetworks could develop pheromones or messages to trigger certain animals behaviors. Therefore, controlling the location of certain animal species in particular environment would be possible. 
        
        \item\textbf {Air pollution control} 
        The quality of air can be managed and controlled by nanonetworks. Advanced nanofilters will be developed  to remove harmful substances or chimicals in the air which will improve the air quality \cite{han2008molecular}. Also, the nanofilters can be used for water quality \cite{shanmuganathan2015experimental}.
        
\end{itemize}

%

\bibliographystyle{IEEEtran}
\bibliography{Security_Project}

\end{document}